\documentclass
{mn2e}
\usepackage{epsfig}
\usepackage{amsmath, amssymb,bm}

\def \be{\begin{equation}}
\def \ee{\end{equation}}

\def \msun{\rm M_{\odot}}

\begin{document}
\title[Why QPEs only in low--mass galaxies?] {Why Are Quasiperiodic Eruptions Only Found in Low--Mass Galaxies?}

\author[Andrew King] 
{\parbox{5in}{Andrew King$^{1, 2, 3}$ 
}
\vspace{0.1in} \\ $^1$ School of Physics \& Astronomy, University
of Leicester, Leicester LE1 7RH UK\\ 
$^2$ Astronomical Institute Anton Pannekoek, University of Amsterdam, Science Park 904, NL-1098 XH Amsterdam, The Netherlands \\
$^{3}$ Leiden Observatory, Leiden University, Niels Bohrweg 2, NL-2333 CA Leiden, The Netherlands}

\maketitle

\begin{abstract}
I consider the current sample of galaxy nuclei producing quasiperiodic eruptions (QPEs). If the quasiperiod results from the orbital motion of a star around the central black hole, the dearth of associated black hole masses $\gtrsim 10^6\msun$ places tight constraints on models. It disfavours those
assuming wide orbits and small eccentricities, because there is ample volume within pericentre to allow significantly more massive holes in QPE systems than are currently observed.
If instead the orbiting star is assumed to pass close to the black hole, the same lack of large black hole masses strongly suggests that the stellar orbits must be
significantly eccentric, with $1 - e \lesssim {\rm few}\times 10^{-2}$. This favours a tidal disruption near--miss picture where QPEs result from repeated accretion from an orbiting star (in practice a white dwarf) losing orbital angular momentum to gravitational radiation, even though this is not assumed
in deriving the eccentricity constraint. Given the tight constraints resulting from the current small observed sample, attempts to find QPE systems in more massive galaxies are clearly important.

\end{abstract}


\footnotetext[1]{E-mail: ark@astro.le.ac.uk}
\section{Introduction}
\label{intro}
Recent observations (Arcodia et al., 2021; Chakraborty et al., 2021; Farrell et al., 2009; Giustini et al, 2020; Miniutti et al., 2019, 2023a, b; Payne et al., 2021) show that several low--mass galaxy nuclei produce quasiperiodic eruptions (QPEs) in ultrasoft X--rays. These systems emit much of their luminosity in short but very bright bursts, and these eruptions recur at slightly varying intervals (see Table 1). 

There are a number of suggested physical models for QPEs (see Table 2). In most cases these explain the quasiperiodicity as the effect of a star orbiting the central massive black hole (MBH) of the galaxy, although a smaller class instead invokes instabilities in an accretion disc around the MBH. Tests of all these models often concentrate on the best--studied source GSN 069. The aim of this paper is to ask what we can learn  
by considering the properties of the whole sample given in Table 1.

Table 1 reveals a clear feature of the current QPE sample: the MBH mass $M_1$ is always fairly small. The largest is $M_1 \simeq 7\times 10^7\msun$ in ASASSN-14ko (see the discussions in King, 2022a,b), and even this is a marked outlier from the  range $1 \lesssim m_5 \lesssim 10 $ of all the others, where $m_5 = M_1/\msun$. This raises an obvious question --
why are there no QPE systems with more massive black holes?
\section{The Black Hole Mass Constraint}
The significance of this black hole mass constraint for any QPE model differs sharply depending on whether this assumes the stellar orbit to be 
distant (i.e. $a \gg R_g$) or close ($a \sim R_g$),
where $a$ is the semi--major axis of the orbit, and $R_g = GM_1/c^2$ the gravitational radius of the MBH.

\subsection{Distant Orbiters}

For distant orbiters the star is very far from filling its tidal lobe throughout its orbit, and the QPE is not powered by direct accretion from it. In particular the orbit cannot have extreme eccentricity $e \gtrsim (1 - R_g/a)$ which would bring it close to the black hole at pericentre. An obvious example of a distant orbiter is a star producing  QPEs by impacting an accretion disc (fuelled in some other way)
around the central black hole twice per orbit. Several variants of this idea have appeared in the literature (see the papers labelled `distant' in Table 2). In all distant orbiter cases the QPE luminosity is limited by the local release of orbital binding energy, and the stellar orbit presumably shrinks slightly during the interaction powering the QPE.\footnote{This suggests that distant orbiters may eventually lose enough angular momentum to evolve into close orbiters, and so ultimately accrete to the black hole. 
} 

For distant orbiters we have $a$ significantly larger than $R_g$. Kepler's law
\be
4\pi^2 a^3 = GM_1P^2 
\label{kep}
\ee
implies that
\be
\frac{a}{R_g} = \left(\frac{cP}{2\pi R_g}\right)^{2/3}\simeq \left(\frac{3P_4}{M_8}\right)^{2/3},
\ee
where $P_4$ is the quasiperiod in units of $10^4$~s and $M_8 = M_1/10^8\msun$,
so that distant orbiters require $M_8 \lesssim P_4$, or 
\be 
M_1 \lesssim 1\times 10^8\msun P_4.
\label{mass}
\ee

We see from Table 1 that this constraint is easily satisfied in all observed QPE systems -- indeed, so easily, that it offers no answer to the question posed in the Introduction. Equation (\ref{mass}) would allow QPE systems with considerably larger black hole masses than actually observed, because there is ample room inside the stellar orbit for them. Every distant orbiter leaves a large volume available for a massive black hole within its orbit.



A possible route out of this difficulty for distant orbiter models would open if one could arrange that  the modelled QPEs had inherently limited luminosities which did not increase significantly with black hole mass.  
Then since  more massive black holes generally have higher accretion luminosities, it would be hard to pick out the relatively faint QPEs in them. The difficulty here is that in all likely models of recurring eruptions, whether involving an orbiting star or a disc instability, the energy release
 ultimately occurs because matter becomes more tightly bound to the central MBH. But this must also be the source of the non--eruptive gas accretion luminosity, suggesting that the ratio of QPE to `steady' flux should be relatively independent of black hole mass (note that all observed QPEs are significantly sub--Eddington for their black hole masses).

\subsection{Close Orbiters}

For close orbiters there is a straightforward reason for an upper limit to the MBH
mass. 
As noted by several authors (e.g. King, 2020), in this case QPEs are closely related to tidal disruption events (TDEs). There, a star falling towards the MBH on a near--parabolic orbit fills its tidal lobe and is torn apart by tidal forces on a dynamical timescale. TDE occurrence significantly favours black holes of lower mass, i.e. $\lesssim 10^7 - 10^8\msun$ (e.g. Kesden, 2012), since for more massive ones the black hole's ISCO radius exceeds the orbital pericentre separation and the star is `silently' swallowed whole by the MBH without filling its tidal lobe.  

In this picture QPEs are TDE near--misses. To give a closed eccentric stellar orbit, the capture event must disrupt a stellar 
binary system (the Hills (1988) mechanism -- see the discussion in Cufari et al. (2022)). Instead of a single star, it is the binary system which is torn apart near the black hole. Its lighter member is ejected with high velocity, carrying off much of the gravitational binding energy of the binary and leaving its more massive star bound to the hole in a highly eccentric orbit. To produce a QPE system, this orbit must remain outside the hole's ISCO radius
even at pericentre, i.e.  
\be 
a(1-e) > \lambda R_g,
\label{tidal}
\ee
but be close enough to the hole at this point to overflow its tidal lobe and feed gas to the black hole (see the papers labelled `close' in Table 2).
Here $\lambda$ is a dimensionless parameter in the range $1 < \lambda < 9$, the limits corresponding respectively to prograde and retrograde orbits around a maximally spinning black hole. Using Kepler's law (\ref{kep}),
eqn (\ref{tidal})
gives a maximum allowed central black hole mass for QPE--producing galaxies
\be
M_1 < M_{\rm max} =  \frac{c^3P}{2\pi G}\left(\frac{1-e}{\lambda}\right)^{3/2}
= 3\times 10^8\msun P_4
\left(\frac{1-e}{\lambda}\right)^{3/2},
\label{tidal2}
\ee
where $P_4$ is the quasiperiod in units of $10^4$~s. 

It is clear that without a significant orbital eccentricity (i.e. $1- e \ll 1$)
the maximum mass allowed by eqn (\ref{tidal2}) would be much larger than seen in any QPE source. But the origin of close orbiters as TDE near--misses makes such extreme eccentricities very reasonable. To rule out
black hole masses $M_1 \gtrsim 10^6\msun$ (i.e. larger than in all systems other than the outlier ASASSN-14ko), eqn (\ref{tidal2}) requires a minimum eccentricity given by
\be 
1 - e \lesssim 0.02\lambda P_4^{-2/3} \lesssim 0.1P_4^{-2/3},
\label{Mmax}
\ee
%
where I have assumed $\langle{\lambda}\rangle < 5$, as prograde and retrograde cases must presumably occur with equal probability, and individual holes may not spin maximally\footnote{From eqn. (\ref{tidal2}), interpreting the low required BH mass as favouring retrograde accretion on to 
maximally spinning holes, i.e. fixing $\lambda = 9$, would still allow BH masses $\gtrsim 10^7\msun$, even if there was a plausible reason for this strange restriction.}.
The more extreme QPE system ASASSN-14ko (where $M_1 \simeq 7\times 10^7\msun$ and $P_4 = 937$) gives a tighter limit
\be
1 - e \lesssim 1.9 \times 10^{-2}.
\label{ecc}
\ee
We shall see below that there are good reasons for these extreme eccentricities.


\section{Discussion}

The considerations of the last Section imply two broad conclusions

1. Distant orbiter models do not currently offer a physical reason for the lack of QPE sources with high mass black holes, and a majority of these systems require some different model. A possible escape route from this conclusion would open if one could show that the modelled QPE events (e.g. disc impacts) themselves had luminosities which did not increase significantly with the black hole mass. Then it might be difficult to detect a QPE signal for more massive black holes with presumably higher `steady' accretion rates.

2. Close orbiter models give a natural explanation for the lack of high black hole masses among QPE sources. Whatever the nature of the
orbiting star, its orbit must be very eccentric (cf eqns \ref{Mmax}, \ref{ecc}).
This requirement  is very natural given the tight relation of the close orbiter picture to the standard model for tidal disruption events, and
follows solely from the observed dearth of QPE black hole masses $\gtrsim {\rm few}\times 10^7\msun$. The eccentricity limit does not rely on any assumption about what drives the mass accretion powering QPE emission, and in particular there is no assumption that this occurs through gravitational radiation losses. If we do assume any particular form of angular momentum loss, the problem becomes determinate, and must give values of $1 - e$ compatible with the limits (\ref{Mmax}, \ref{ecc}) -- see Table 1 of King, 2023, where driving by gravitational radiation is assumed, and which in particular shows that ASASSN-14ko must have an eccentricity significantly more extreme than required by the limit (\ref{ecc}).

It is clear that the required very high eccentricities, together with the observed short periods (Table 1) do strongly favour models where loss of the orbiting star's angular momentum 
via gravitational radiation drives the accretion powering QPEs. But this was {\it not} assumed in deriving the eccentricity limits. 

In practice (see King, 2020, 2022) the resulting QPEs attain the observed luminosities  only if the close orbiter is a white dwarf (MS) rather than a more extended (e.g. main--sequence) star, which must give a longer period and so weaker gravitational radiation and so smaller accretion rates.
This type of closely orbiting WD model gives modest values for their masses for the known QPE sample (see King, 2022, 2023b). There is also a direct observation of the
CNO enhancement likely for a WD companion in at least one case (Sheng, 2021).  


Escape from these rather tight constraints might be possible if future observations reveal a population of QPE sources with significantly more massive black holes. As the current sample is still very small, searches for systems like this are clearly warranted. 


\section*{DATA AVAILABILITY}
No new data were generated or analysed in support of this research.
\section*{ACKNOWLEDGMENTS}
I particularly thank Giovanni Miniutti for cordial and continuing discussions on many aspects of QPE sources.

\begin{table*}
\caption{Parameters of the Current QPE Sample}
\vskip 5.0pt
\centering
{
\setlength{\tabcolsep}{3pt}
{
\hfill{}
  \begin{tabular}{|l||c|c|c|} 
    \hline
    Source & $P_4$ & $m_5$  
      \\
    \hline \hline
    \\
    eRO--QPE2$^a$ & 0.86&  2.5 
    \\
  XMMSL1 
J024916.6--04124$^b$ & 0.90 &  0.85   
    \\
    RXJ1301.9+2747$^c$ &1.65& 18   
    \\
    GSN 069$^d$ & 3.16&  4.0 
    \\ 
     eRO--QPE1$^e$ & 6.66 &9.1
     \\
     ASASSN--14ko$^f$ &937& 700
     \\
     HLX--1$^g$ & 2000& [10] 
     \\
     \  . . . & \ . . .  & [0.5]
     \\
   \hline\hline
  \end{tabular}}
  \hfill{}
  }
  \vskip 0.2truecm
  \begin{itemize}
\item[] 
$P_4$ is the observed quasiperiod in units of $10^4$~s, and $m_5$ the MBH mass in units of $10^5\msun$. 
References to observations: 
$a =$ Arcodia et al. (2021),
$b = $ Chakraborty et al. (2021), 
$c = $ Giustini et al. (2020),
$d = $ Miniutti et al (2019),
$e = $ Arcodia et al. (2021),
$f = $ Payne et al. (2021),
$g = $ Farrell et al. (2009). Note that there is no secure mass for the black hole in HLX--1, and $m_5 = 0.5$ is the minimum allowing a WD mass respecting the Chandraskhar limit. Assuming a larger $m_5$ gives smaller $m_2$ and $e$.

\end{itemize}
\end{table*}


\begin{table*}
\caption{QPE Models}
\vskip 5.0pt
\centering
{
\setlength{\tabcolsep}{3pt}
{
\hfill{}
  \begin{tabular}{|l|c|c|c|} 
    \hline
    reference & close/distant orbiter & eccentricity range
      \\
    \hline \hline
    Chen et al., 2022 & close &0.7
    \\
    Franchini et al., 2023 & distant & $0.05 - 0.5$
    \\
    King, 2020, 2022, 2023a, 2023b & close & $\gtrsim 0.95$
    \\
    Krolik \& Linial 2022 & distant & $0.1 - 0.5$
    \\
    Linial \& Metzger, 2023 & distant & $\gtrsim 0.5$
    \\
    Linial \& Sari, 2023 & distant & $0.1 - 0.2$
    \\
    Lu \& Quataert, 2022 & distant & $\sim 0.5$
    \\
    Metzger et al. 2022 & distant & $ \ll 1$
    \\
    Tagawa \& Haiman, 2023 & distant & 0
    \\
    Sukov\'a et al., 2021& close & $\gtrsim 0.95$
    \\
    Wang et al., 2022 & close & $\gtrsim 0.95$
    \\
     Xian et al., 2021 & distant & $\sim 0.05$
     \\
     Zhou et al., 2022 &  close &  $\gtrsim 0.95$ 
     \\
       \hline\hline
  \end{tabular}}
  \hfill{}
  }
  \vskip 0.2truecm
  \begin{itemize}
\item[] 

\end{itemize}
\end{table*}

{}


\begin{thebibliography}{}

\bibitem{} 
Arcodia, R., Merloni, A., Nandra, K., et al., 2021, Nature, 592, 704


\bibitem{} 
Chakraborty, J., Kara, E., Masterson, M., et al., 2021, ApJL, 921, L40 

\bibitem{}
Chen, X., Qiu,Y., Li, S., Liu, F.K. 2022, ApJ, 930, 122

\bibitem{}
Cufari, M., Coughlin, E.R., Nixon, C.J., 2022, ApJ, 929, L20
\bibitem{}
Farrell S. A., Webb N. A., Barret D., Godet O., Rodrigues J. M.,
2009, Nature, 460, 73
\bibitem{}
Franchini, A., Bonetti, M., Lupi, A., 2023
arXiv 230400775

\bibitem{} 
Giustini, M., Miniutti, G., \& Saxton, R. D., 2020, A{\&}A,
636, L2 
\bibitem{}
Kesden, M., 2012, Physical Review D, vol. 85, Issue 2, id. 024037
\bibitem{}
Hills, J. G. 1988, Nature, 331, 687
\bibitem{} 
King, A.R., 2020, MNRAS, 493, L120
\bibitem{}
King, A.R., 2022, MNRAS, 515, 4344
\bibitem{}
King, A.R., 2023a, MNRAS, 520, L63K
\bibitem{}
King, A.R., 2023b, MNRAS, 523, L26K
\bibitem{}
Krolik, J., Linial., I., 2022, ApJ, 941, 24
\bibitem{}
Linial, I., Metzger, B.D., 2023, arXiv 23031623
\bibitem{}
Linial, I., \& Sari, R., 2017, MNRAS 469, 2441L 





\bibitem{}
Lu, Wenbin \& Quataert, E., 2022, arXiv 221008023  
\bibitem{}
Metzger, B.D., Stone, N.C., Gilbaum, S., 2022, ApJ 926, 101
\bibitem{}
Miniutti, G.; Giustini, M.; Arcodia, R. et al., 2023a,
arXiv 230509717
\bibitem{}
Miniutti, G.; Giustini, M.; Arcodia, R. et al., 2023b,
A\&A 670, 93
\bibitem{} 
Miniutti, G., Saxton, R. D., Giustini, M. et al., 2019,
Nature, 573, 381 
\bibitem{}
Pan Xin, Li, Shuang-Liang; Cao, Xinwu et al., 2022, ApJ, 928L 18
\bibitem{}
Payne, A. V., Shappee, B. J., Hinkle, J. T., et al. 2021, ApJ, 910, 125




\bibitem{}
Sheng, Z.,  Wang, T.,  Ferland, G., et al.,  
2021, ApJ 920L, 25
\bibitem{}
Sukov\'a, P., Zajacek, M., Witzany, V. et al., 2021, ApJ, 917, 43
\bibitem{}
Sun, L., Shu, X., Wang, T., 2013, ApJ 768, 167
\bibitem{}
Tagawa, H., Haiman, Z., 2023, arXiv 2304.03670



\bibitem{}
Wang, M., Yin, J., Ma, Y., Wu, Q., 2022, ApJ, 933, 225
\bibitem{}
Xian, J., Zhang, F., Dou, L., et al.,
2021, ApJ 921L, 32
\bibitem{}
Zhao, Z. Y., Wang, Y. Y., Zou, Y. C., Wang, F. Y., Dai, Z. G.,
2022, A\&A, 661, A55



\bsp
\end{thebibliography}
\end{document}